\documentclass[11pt]{article}
\usepackage[english]{babel}
\usepackage{amssymb,amsfonts,amsmath,cite,enumerate,float,cmap,mathrsfs,indentfirst,graphicx,caption2,subfigure,geometry,titlesec,setspace}
\graphicspath{{pictures/}}
\DeclareGraphicsExtensions{.png}
\usepackage{slashed}
\usepackage{mathtools}
\usepackage{simplewick}
\usepackage{simpler-wick}
\usepackage{amsxtra}
\usepackage[export]{adjustbox}
\usepackage{hyperref}
\usepackage{authblk}
\usepackage{tikz}

\title{Local temperature distribution in the vicinity of gravitationally bound objects in the expanding Universe}
\author{P. M. Petryakova \\ petryakova.pm@ya.ru \\ S. G. Rubin \\ sergeirubin@list.ru }
\date{October 2019}
\affil{National Research Nuclear University MEPhI (Moscow Engineering Physics Institute), 115409, Kashirskoe shosse 31, Moscow, Russia}
\begin{document}
\maketitle 
\begin{abstract}
We consider a cluster of Primordial Black Holes which is decoupled from the cosmological expansion (Hubble flow) and this region is heated as compared to the surrounding matter. The increased temperature inside the region can be explained by several mechanisms of Primordial Black Holes formation. We study the temperature dynamics of the heated region of Primordial Black Holes cluster.
\end{abstract}
\section*{Introduction}
\par The idea of the Primordial Black Holes (PBH) formation was predicted five decades ago \cite{1}. Although they have not yet been identified in observations but some astrophysical effects can be attributed to PBH: supermassive black holes in early quasars. Therefore till now, PBH give information about processes in the Early Universe only in the form of restrictions on the primordial perturbations \cite{5} and on physical conditions at different epochs. It is important now to describe and develop in detail models of PBH formation and their possible effects in cosmology and astrophysics.
\par There are several models of PBH formation. PBH can be formed during the collapses of adiabatic (curvature) density perturbations in relativistic fluid \cite{20}. They could be formed as well at the early dust-like stages \cite{21} and rather effectively on stages of dominance of dissipative superheavy metastable particles owing to a rapid evolution of star-like objects that such particles form \cite{23}. There is also an exciting model of PBH formation from the baryon charge fluctuations \cite{25}.
Another set of models uses the mechanism of domain walls formation and evolution with the subsequent collapse \cite{27}. Quantum fluctuations of a scalar field near a potential maximum or saddle point during inflation lead to the formation of closed domain walls \cite{PBH}. After the inflation is finished, the walls could collapse into black holes in the final state. There is a substantial amount of the inflationary models containing a potential of appropriate shape. The most known examples are the natural inflation \cite{30} and the hybrid inflation \cite{31} (and its supergravity realization \cite{33}). The landscape string theory provides us with a wide class of the potentials with saddle points, see review \cite{34} and references within. Heating of the surrounding matter is the inherent property of the domain wall mechanism of PBH cluster formation. While collapsing the domain wall partially transfers its kinetic energy to the ambient matter. It would allow to distinguish different models by observations.

\section{The first Chapman{--}Enskog approximation}

According to the discussion above, PBH are gathering into the clusters with heated media inside them. 
It is assumed that after decoupling from the cosmological expansion the temperature of gas inside the cluster and its density is higher than that around the cluster. These factors can ignite a new chain of nuclear reactions changing chemical composition of the matter in given region. We are going to study the rate of temperature spreading into surrounding space and the temperature distribution within the cluster. The temperature dynamics is described by the appropriate equations in the framework of the Chapman{--}Enskog procedure.

 The Chapman{--}Enskog method \cite{CH} makes it possible to obtain a solution to the transport equation and it can be applied to the relativistic transport equation in general case. The applicability condition of this method: macroscopic wavelengths should be significantly greater than the mean free path. This excludes the propagation velocity that is faster than the thermal velocity of particles \cite{W}. Using this method, we can find linear laws for flows, thermodynamic forces and expressions for transfer coefficients based on the solution of linearized transfer equation. After that we apply this linear laws to continuity, energy and motion equations. This leads to the relativistic Navier{--}Stokes equations which form a closed system for hydrodynamic variables. In the first approximation various irreversible flows are linearly related to non-uniformities present in the system. In this case the relativistic generalization of the Fourier{--}law for the heat flux and the linear expression for the viscous pressure tensor has the form ($c=\hbar=k_B=1$)
\begin{align}\label{Fouurier}
I^{\mu}_{q}&=\lambda\Bigl(\nabla^{\mu}T-\frac{T}{hn}\nabla^{\mu}\text{p}\Bigr) \\\label{Pi}
\Pi^{\mu\nu}&=2\eta\mathring{\overline{\nabla^{\mu}u^{\nu}}}+\eta_{v}\Delta^{\mu\nu}\nabla_{\sigma}u^{\sigma}
\end{align}
 $\lambda$ {--} the heat conductivity, $\eta$ {--} the shear viscosity, $\eta_{v}$ {--} the volume viscosity, $\nabla^{\mu}=\Delta^{\mu\nu}\partial_{\nu}$ , $\Delta^{\mu\nu}=g^{\mu\nu}-u^{\mu}u^{\nu}$ and this operator acts as a projector: $\Delta^{\mu\nu}u_{\nu}=0$. They are designed to select two hydrodynamic four{--}velocity expressions proposed by Eckart and Landau{--}Lifshitz. We will use the definition of Eckart 
 which relates the hydrodynamic four{--}velocity directly to the particle four{--}flow $N^{\mu}$
\begin{equation}
    u^{\mu}=\cfrac{N^{\mu}}{\sqrt{N^{\nu}N_{\nu}}} \,.
\end{equation}
The relativistic equation of motion and equation of energy are given by \cite{Gr}
\begin{align}\label{eqmot}
hnDu^{\mu}=\nabla^{\mu}\text{p}-&\Delta^{\mu}_{\nu}\nabla_{\sigma}\Pi^{\nu\sigma} + (hn)^{-1}\Pi^{\mu\nu}\nabla_{\nu}\text{p}- \\ \nonumber -\bigl(&\Delta^{\mu}_{\nu}DI^{\nu}_{q}+I^{\mu}_{q}\nabla_{\nu}u^{\nu}+I^{\nu}_{q}\nabla_{\nu}u^{\mu}\bigr) 
\end{align}
\begin{equation}
\label{energy}
nDe=-\text{p}\nabla_{\mu}u^{\mu}+\Pi^{\mu\nu}\nabla_{\nu}u_{\mu}-\nabla_{\mu}I^{\mu}_{q}+2I^{\mu}_{q}Du_{\mu}.
\end{equation}
After linearization, the energy equation is reduced to 
\begin{equation}\label{Energy}
\cfrac{DT}{T}=\cfrac{1}{c_{v}}\Biggl[\nabla_{\mu}u^{\mu}-\frac{\lambda}{\text{p}}\Bigl(\nabla^{2}T-\frac{T}{hn}\nabla^{2}\text{p}\Bigr)\Biggr]
\end{equation}
where we have taken into account the linear laws \eqref{Fouurier} and \eqref{Pi}, $\nabla^2$ stands for $\nabla^2=\nabla^{\mu}\nabla_{\mu}$ and $D=u^{\mu}\partial_{\mu}$.

If the hydrodynamic four{--}velocity is constant and $ \text{p}=nT$ (we will show it in the next section) the energy equations reduce to the relativistic heat-conduction equation:
\begin{equation}\label{ONO}
 nc_{v}DT=-\lambda\Biggl(\nabla^{2}T-\frac{T}{hn}\nabla^{2}\text{p}\Biggr).
\end{equation}
\section{The thermodynamic values}
\par The equilibrium distribution function with no external fields takes the form of the Juttner distribution function
\begin{equation}
f(p)=\cfrac{1}{\bigl(2\pi \bigr)^3}\,\, \text{exp}\biggl(\cfrac{\mu-p^{\mu}u_{\mu}}{T}\biggr) .
\end{equation}
It allows to calculate the particle four{--}flow in equilibrium
\begin{equation}
    N^{\mu}=\cfrac{1}{\bigl(2\pi \bigr)^3}\, \int \cfrac{d^3p}{p^0} \, p^{\mu} \, \text{exp}\biggl(\cfrac{\mu-p^{\mu}u_{\mu}}{T}\biggr) \, .
\end{equation}
The Juttner distribution function outlines one direction in space{--}time. As a result, it must be proportional to four{--}velocity, where the proportionally factor of this relation is the particle density
\begin{equation}
    n=\cfrac{1}{\bigl(2\pi \bigr)^3}\, \int \cfrac{d^3p}{p^0}\, p^{\mu} \, u_{\mu}\, \text{exp}\biggl(\cfrac{\mu-p^{\nu}u_{\nu}}{T}\biggr) \, .
\end{equation}
The integral is a scalar and it can be calculated at selected $u^{\mu}=(1,0,0,0)$. This result can be expressed in the modified Bessel function of the second kind
\begin{equation}
    n=\cfrac{4 \pi m^2 T}{\bigl(2\pi \bigr)^3}\, K_2 \biggl(\cfrac{m}{T}\biggr) \text{exp}\biggl(\cfrac{\mu}{T}\biggr) \, .
\end{equation}
We can obtain the equilibrium pressure following the same reasoning to calculate the energy{--}momentum tensor in equilibrium:
\begin{align}\label{pnT}
    \text{p}&= -\cfrac{1}{3}T^{\mu\nu}\Delta_{\mu\nu}= -\cfrac{1}{3} \int \cfrac{d^3p}{p^0}\, p^{\mu} \, p^{\nu}\,\Delta_{\mu\nu}f(p) = \\ \nonumber &=\cfrac{4 \pi m^2 T^2}{\bigl(2\pi \bigr)^3}\, K_2 \biggl(\cfrac{m}{T}\biggr) \text{exp}\biggl(\cfrac{\mu}{T}\biggr)=nT \, .
\end{align}
Hence, if we identify T with the temperature of the system the standard scheme of thermodynamics could be clearly seen.

 Using recurrence relation for the modified Bessel function of the second kind and taking into account particle density the expression has the form
\begin{equation}\label{e}
    e=m\,\dfrac{K_{3}(m/T)}{K_{2}(m/T)}-T \, .
\end{equation}
Considering the result of \eqref{pnT} for pressure we can find the enthalpy per particle
\begin{equation}\label{h1}
    h=e+pn^{-1}=m\, \dfrac{K_{3}(m/T)}{K_{2}(m/T)}-T+T=m\,\dfrac{K_{3}(m/T)}{K_{2}(m/T)} \, .
\end{equation}
The heat capacity per particle at constant volume by definition
\begin{equation}\label{cv}
    c_{v}=\partial e/ \partial T \, .
\end{equation}
 We can get asymptotic behaviour of these values for large arguments of the modified Bessel function of the second kind (which corresponds to the case of low temperatures) and for small arguments (which corresponds to the case of massless particles). For small values of temperature we have the asymptotic ratio for large arguments ($w=m/T$):
\begin{equation}\label{asimptK}
    K_{n}(w)\simeq \cfrac{1}{e^{w}} \, \sqrt{\frac{\pi}{2w}} \, \Biggl[1+ \cfrac{4n^{2}-1}{8w}+\cfrac{(4n^{2}-1)(4n^{2}-9)}{2!(8w)^{2}}+ \ldots \, \Biggr] \, .
\end{equation}
It allows to obtain the enthalpy per particle:
\begin{equation}\label{h}
    h=m+\cfrac{5}{2}\,T+\cfrac{15}{8}\, \cfrac{T^{2}}{m}+ \ldots  
\end{equation}
and to derive the caloric equation of state of relativistic perfect gas \eqref{e} and the heat capacity per particle at constant volume \eqref{cv}:
\begin{equation}
  e=m+\cfrac{3}{2}\,T+\cfrac{15}{8}\, \cfrac{T^{2}}{m}+ \ldots  
\end{equation}
\begin{equation}\label{cvv}
     c_{v}=\cfrac{\partial e}{\partial T}= \cfrac{3}{2}\,+\cfrac{15}{4}\, \cfrac{T}{m}+ \ldots  
\end{equation}

Massless particles are essential in relativistic kinetic theory. For this purpose we should expand our formulas in this special case. The results can be obtained by taking the limit in $m\rightarrow 0$ with the asymptotic relation for the modified Bessel function of the second kind:
\begin{equation}
    \lim_{w\to 0}  w^{n} K_{n}(w)=2^{n-1}(n-1)!
\end{equation}
\begin{equation}\label{Thigh}
    e=3T \,\, , \quad h=4T \,\, , \quad c_{v}=3 \, .
\end{equation}
In this case, with the caloric equation of state of relativistic gas and $\text{p}=nT$ we can obtain well-known expression for pressure for massless particles: $\text{p}=en/3$.

We can find Fourier differential equation of the heat conduction in the non{--}relativistic case. Following expression \eqref{h} in the case of low temperatures ($ T \ll m$) and considering the ratio $ \text{p}=nT $ we obtain
\begin{equation}\label{nonrel}
 nc_{v}DT=-\lambda\Bigl(\nabla^{2}T-\cfrac{T}{hn}\nabla^{2}\text{p}\Bigr)\eqsim-\lambda\nabla^{2}T.
\end{equation}
In this case, the heat{--}conduction equation allows an infinite propagation
velocity. Although this feature is already present in the non{--}relativistic theory in the relativistic theory it becomes a paradox: the thermal disturbances can not propagate faster than the speed of light. This paradox is easily resolved in the framework of the Chapman{--}Enskog procedure. In fact the restriction inherent in the Chapman{--}Enskog method (the macroscopic wave lengths has to be much greater than the mean free path) prevents the existence of propagation velocities faster than the thermal velocity of particles.

\section{Thermal equilibrium}
\par We should check the applicability of our results by estimating to what extent the electron{--}proton{--}photon plasma is close to kinetic equilibrium before and during recombination. All our previous calculations were made under the assumption that the distribution functions have equilibrium form and all components have the same temperature equal to the photon temperature. To make sure that the temperature of electron{--}proton component coincides with the photon temperature we have to study the following effect. The effective temperature of photons would decrease in time slower than that of electrons and protons. Thus we have to check that energy transfer from photons to electrons and protons is sufficiently fast.
\par Electrons get energy from photons via Compton scattering process that occurs with Thomson cross section. The time between two subsequent collisions of a given electron with photons is
\begin{equation}
    \tau =\cfrac{1}{n_{\gamma}\sigma_{T}}
\end{equation}
here $\sigma_T$ {--} the Compton cross section and $n_\gamma$ {--} the number density of photons. For energy transfer the time $\tau_E$ in which an electron obtains kinetic energy of the same order of magnitude as temperature due to the Compton scattering should be found. We note that the typical energy transfer in a collision of a slow electron with a low energy photon is actually suppressed for estimation of this time. The estimation for number of scattering events needed to heat up a moving electron is given by \cite{Rubakov}
\begin{equation}
    N \sim \biggl(\cfrac{T}{\Delta E}\biggr)^2 \sim \cfrac{m_e}{T} \, .
\end{equation}
We can obtain the time of electron heating 
\begin{equation}
    \tau_E(T)\sim N\tau(T) \sim \cfrac{m_e}{n_{\gamma}(T)\sigma_T T} \, .
\end{equation}
At the moment of recombination $ \tau_E(T_{rec}) \simeq 6$ yrs. It is much smaller than the Hubble time and energy transfer from photons to electrons is efficient. Thus electrons and protons have the photon temperature.
\par What about the heating of protons? Doing the same procedure $\bigl($with $m_p$ substituted for $m_e$ in (25)$\bigr)$ we obtain that process of direct interaction of proton with photons is irrelevant. Since the Thomson cross section is proportional to $ m_{e}^{-2}$ the time for protons is larger by a factor $\bigl(m_p/m_e\bigr)^3$ and this time is larger than the Hubble time. Energy transfer to protons occurs due to elastic scattering of electrons off protons. The energy transfer time 
is
\begin{equation}\label{te}
    \tau_E(T)\sim \cfrac{m_e m_p}{16 \pi n_{e}(T) \alpha^2 \ln (6Tr_D/\alpha)} \biggl(\cfrac{3T}{m_e}\biggr)^{3/2}
\end{equation}
here $r_D =\sqrt{\cfrac{T}{4 \pi n_e \alpha}} $ and during recombination  $\tau_E(T_{rec}) \sim 10^4 $ s and this time is very small compared to the Hubble time at recombination. The estimation done for protons is valid for electrons as well with $m_e$ substituted for $m_p$ and numeric factor. This means that electrons and protons have equilibrium distribution functions with temperature equal to photon temperature.
\section{Transport coefficients}
The divergence of the collision integrals is the main difficulty encountered when applying the transport equation to plasma. The many particle correlations which provide the Debye shielding are not included in the transport equation due to the long range nature of electromagnetic interaction. In the Standard Model of the Universe Compton scattering between photons and electrons was the dominant mechanism for energy and momentum transfer in the radiation{--}dominated era (RD{--}stage). It seems worthy to present a quantitative description of the non{--}equilibrium processes that can be expected in a hot photon gas coupled to plasma by Compton scattering.

In case of low temperatures we have the following expression for heat conductivity 
\begin{equation}
    \lambda=\cfrac{4}{5}\cfrac{x_\gamma}{x_e}\cfrac{1}{\sigma_{T}}
\end{equation}
here $x_i$ {--} the fraction of particles, $\sigma_{T}$ {--} the Compton cross section and the ratio of electron and photon number densities through baryon{--}to{--}photon ratio with electric neutrality of the Universe
\begin{equation}
    \eta_B = \cfrac{n_B}{n_\gamma} =0.6 \, 10^{-9} \, .
\end{equation}
\section{Dependence of the equation on the rate of \,\,\, expansion of space}
We should set the form of operators included in the equation \eqref{ONO}. If the matter of the surrounding space is stationary as a whole then the four{--}velocity takes the form $ u_\mu=(1,0,0,0) $ hence $ D=u^\mu \partial_\mu= \partial_t $. We need to make the following substitution: $\nabla^{\mu}=\Delta^{\mu\nu}\partial_{\nu} \longrightarrow \Delta^{\mu\nu}\mathscr{D}_{\nu}$ in order to take into account the expansion of space. 
Thus our operator $\nabla^2$ is explicitly dependent on the metric
\begin{align}\label{nablaa}
\nabla^{2}&=\nabla^\mu \nabla_\mu=\nabla^\mu g_{\mu\nu}\nabla^\nu=\Delta^{\mu\nu}\mathscr{D}_{\nu}g_{\mu\nu}\Delta^{\nu k}\partial_{k} \, .
\end{align}
The rate of temperature spreading into the surrounding space will be calculated with respect to the Friedmann{--}Lemaître{--}Robertson{--}Walker metric. The metric tensor in this case has the form
\begin{equation}
    g_{\mu\nu}=\text{diag}(1,-a^2(t),-a^2(t)r^2,-a^2(t)r^2\sin^2\theta) \, .
\end{equation}
The scale factor $a(t)$ can be found from Friedmann equations.
Finally we get the following dependence for scale factor obtained under the conditions $ a(t_{0})=1 $, $\rho(t_{0})= 0.53\cdot 10^{-5} \,\text{GeV}/\text{cm}^{3}$, $t_{0}\simeq 14\cdot 10^9\, \text{yrs}${--} the age of Universe
\begin{equation}\label{at}
  a(t)=\biggl[1+\cfrac{3\xi}{2}\sqrt{\frac{8\pi G \rho_{0}}{3}}(t-t_{0})\biggr]^{2/3\xi} \, ,
\end{equation} 
 here $\xi=4/3 $ for stage of radiation dominance (RD{--}stage) and  $\xi=1$  for stage of the matter domination (MD{--}stage).
\section{Final statement of the problem and result of calculation}
We consider spherical symmetry for simplicity. The heat{--}conduction equation \eqref{ONO} with expression \eqref{pnT} for pressure and in case of stationary matter takes the form
\begin{equation}\label{ONOFINAL}
 \cfrac{nc_v}{\lambda}\cfrac{\partial}{\partial t}\Bigl( T(r,t)\Bigr)=-\nabla^{2}T(r,t)+\frac{T(r,t)}{hn}\nabla^{2}nT(r,t)
\end{equation}
with boundary and initial conditions
\begin{equation}\label{BC}
\begin{cases}
\cfrac{\partial}{\partial r}T(r,t)\bigg|_{r=0}=0 \, ,  \\
T(r,t)|_{r=\infty}=\cfrac{T_{out}}{a(t)} \, ,
\end{cases}
T(r,0)=T_{in}\exp(-r^{2}/r_{0}^{2})+T_{out} \, ,
\end{equation}
here the dependence for scale factor $a(t)$ is taken from \eqref{at}, $T_{in}$ and $T_{out}$ are temperatures of matter inside cluster and the surrounding space respectively, $r_0$ is temperature distribution parameter.

\begin{figure}[h!]
 \begin{center}
   \includegraphics[width=0.25\textwidth]{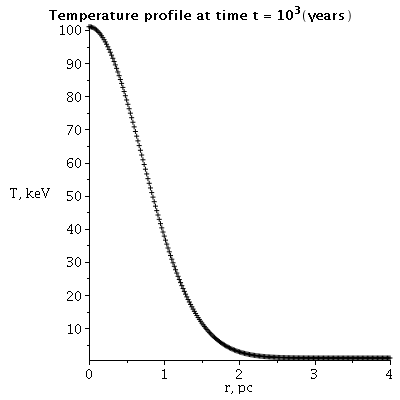}
   \label{fig:T0}
  \hspace{0.12cm}
   \includegraphics[width=0.25\textwidth]{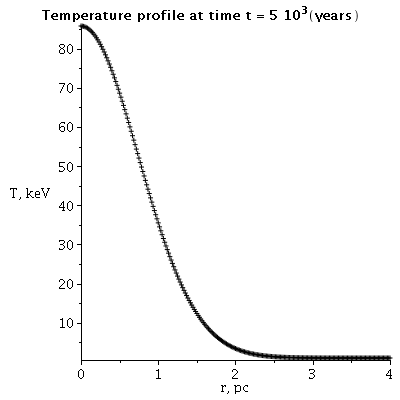}
   \label{fig:T5}
   \includegraphics[width=0.25\textwidth]{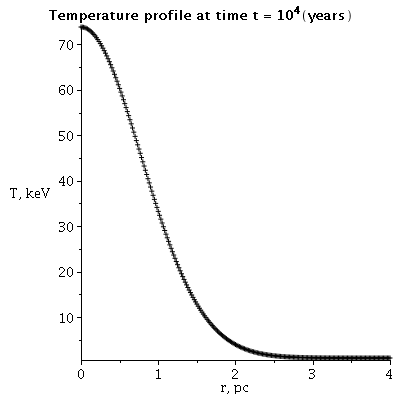}
   \label{fig:T10}
 \end{center}
 \caption{Numerical calculation results of local temperature distribution at the RD{--}stage.}
 \label{fig:solution}
\end{figure}

In general the obtained expressions can also be used in calculations at the RD{--}stage and the MD{--}stage. For this purpose the expression for the scale factor \eqref{at} should be taken at different $ \xi $ and with modified heat conductivity. Presumably the cluster of primordial black holes virializes at the end of the RD{--}stage. It makes sense to estimate its cooling before the end of this stage. We need to choose specific values of the following parameters:

\begin{tabular}{ll}
     & $\cdot$ temperature distribution parameter $r_{0}=1$ pc;\\
	 & $\cdot$ temperature inside the area $T_{in}=100$ keV;\\
	 & $\cdot$ temperature of the surrounding space $T_{out}=1$ keV;\\
	 & $\cdot$ dependence a(t) in boundary condition is selected for RD{--}stage;\\
	 & $\cdot$ for enthalpy and heat capacity we should select forms in case of\\ 
	 & $ \,\, $ low temperatures \eqref{h} and \eqref{cvv} accordingly.
\end{tabular}
Using numerical simulation we have Fig.\ref{fig:solution}. As can be seen from the figure, the gravitationally bound region almost completely retains temperature which was obtained during the formation at the RD{--}stage. The next step is to determine what happens with this heated region at the MD{--}stage.
\section{Estimation for MD{--}stage}

We will be interested in the internal temperature of the gravitationally bound region during the MD{--}stage. At the end of the RD{--}stage we have a region with higher temperature. It is possible to ignite a new chain of nuclear reactions changing chemical composition of the matter in given region. The temperature inside the cluster can be calculated in Minkowski space and we can find the dependence of the thermal conductivity on temperature in the non{--}relativistic case. The thermal diffusivity by definition is given by \footnote{ Here the values are expressed in the CGS system and the temperature in eV} 
\begin{equation}\label{coeffTemp}
  \chi= \cfrac{\lambda}{n_e c_{v}} = 3.16 \, \cfrac{T_{e}\tau_{e}}{m_{e} c_{v}}= \cfrac{3.16}{2\sqrt{2\pi} \, \sqrt{m_{e}} q_{e}^{4}Z n_{e} (T)\ln (6Tr_D/\alpha)}\,T_{e}^{5/2}
\end{equation}
The thermal diffusivity in $\biggl[ \cfrac{\text{pc}^2}{\text{year}}\biggr] $ is 
\begin{equation}
       \chi(T) \simeq \cfrac{2.3 \cdot 10^{-14}}{ \sqrt{T_e(\text{eV}})} \, .
\end{equation}
The calculated value allows to retain the increased temperature inside the cluster until the recombination starts. The heat is conserved within a region starting from the moment of its formation. Thus, there are significant prerequisites for anomalies in the chemical composition of this region which makes sense to consider in future.

\section*{Conclusion}

We investigated the temperature dynamics of the heated region around the primordial black holes cluster.  For this purpose the relativistic heat{--}conduction equation (without convective terms) was considered taking into account the expansion of space in the framework of the Chapman{--}Enskog relativistic procedure.  The numerical solution was found with the corresponding initial and boundary conditions.  According to our calculations, the  gravitationally  bound  region  almost  completely  retains  temperature  which  was obtained  during  the  formation. At the MD{--}stage the increased temperature inside the cluster is conserved until then recombination will start.  Thus, there are significant prerequisites for anomalies in the chemical composition of this region. In prospect, we are going to study possible anomalies in the chemical content of the region with comparison to the observed data.

\section*{Acknowledgement}

The work of S.G.R. is supported by the grant RFBR N~19-02-00930 and is performed according to the Russian Government Program of Competitive Growth of Kazan Federal University. The work was also supported by the Ministry of Education and Science of the Russian Federation, MEPhI Academic Excellence Project (contract N~02.a03.21.0005, 27.08.2013).

\end{document}